# A health concern regarding the protein corona, aggregation and disaggregation

Mojtaba Falahati[1]*, Farnoosh Attar[2], Majid Sharifi[1], Thomas Haertlé[3,4,5], Jean-François Berret[6], Rizwan Hasan Khan[7], and Ali Akbar Saboury[5]

[1]Department of Nanotechnology, Faculty of Advanced Science and Technology, Tehran Medical Sciences, Islamic Azad University, Tehran, Iran.
[2]Department of Biology, Faculty of Food Industry & Agriculture, Standard Research Institute (SRI), Karaj, Iran.
[3]UR1268, Biopolymers Interactions Assemblies, INRA, BP 71627, 44316 Nantes Cedex 3, France
[4]Poznan University of Life Sciences, Department of Animal Nutrition and Feed Management, ul. Wołyńska 33, 60-637 Poznań, Poland
[5]Institute of Biochemistry and Biophysics, University of Tehran, Tehran, Iran
[6]Matière et Systèmes Complexes, UMR 7057 CNRS Université Denis Diderot Paris-VII, Bâtiment Condorcet, 10 rue Alice Domon et Léonie Duquet, F-75205 Paris, France
[7]Molecular Biophysics and Biophysical Chemistry Group, Interdisciplinary Biotechnology Unit, Aligarh Muslim University, Aligarh-202002, India.

**Corresponding Author:** Mojtaba Falahati, Email: falahati@ibb.ut.ac.ir
Tel: +98 21 22640055    Fax: +98 21 22640059

**Abstract**
Nanoparticle (NP)–protein complexes exhibit the "correct identity" of NP in biological media. Therefore, protein–NP interactions should be closely explored to understand and to modulate the nature of NPs in medical implementations. This review focuses mainly on the physicochemical parameters such as dimension, surface chemistry, the morphology of NPs and influence of medium pH on the formation of protein corona and conformational changes of adsorbed proteins by different kinds of methods. Also, the impact of protein corona on the colloidal stability of NPs is discussed. Uncontrolled protein attachment on NPs may bring unwanted impacts such as protein denaturation and aggregation. In contrast, controlled protein adsorption by optimal concentration, size, pH and surface modification of NPs may result in potential implementation of NPs as therapeutic agents especially for disaggregation of amyloid fibrils. Also, the effect of NPs-protein corona on reducing the cytotoxicity and clinical implications such as drug delivery, cancer therapy, imaging and diagnosis will be discussed. Validated correlative physicochemical parameters for NP–protein corona formation frequently derived from protein corona fingerprints of NPs which are more valid than the parameters obtained only on the base of NP features. This review may provide useful information regarding the potency as well as the adverse effects of NPs to predict their behavior in the *in vivo* experiments.

**Abrivation keys:** Bovaine serum albumin (BSA), circular dichroism (CD), cytochrome c (Cyt c), Gold (Au), Human serum albumin (HSA), Mass spectrometry (MS), Nanoparticles (NPs), Reactive oxygen species (ROS), Silver (Ag), Sodium dodecyl sulfate electrophoresis (SDS-PAGE), Superparamagnetic iron oxide NPs (SPIONs), Transmission electron microscopy (TEM).

**Contents**

**1. Introduction**
A basic understanding of physicochemical properties of NP which impact the adsorption of protein is decisive for the design and development of well-characterized nanomaterials. These distinctive nanomaterials may be implemented in a number of areas, from sensing, where they can be employed for the detection of several biomolecules [1-3], to biomedical engineering, where they are utilized for targeted drug delivery to cancerous tissues [4-8]. NPs are very attractive in new healthcare implementations such as vaccinations [9, 10], cancer diagnosis and therapy [11-15]. NPs display distinctive features in the human body, but there is only a limited knowledge of how these NPs interact with cells and biomolecules such as proteins. Depending on the NP dimension, pH of the medium, surface chemistry and type of NPs and proteins, proteins can mediate the formation of a corona at the external surface of NPs [16-18]. The unique protein coronas of differently modified or differently sized NPs affect all features of NP – biological systems interactions ranging from induced cytotoxicity to the mechanism of infiltration [19-23]. NPs after uptaking by a physiological environment are coated with biomolecules almost spontaneously [24, 25].

The long-lived protein or short-lived protein layers adsorbed tightly or weakly on the NP surface are considered as the hardest coronas (HC) or softest coronas (SC), respectively [26, 27]. Hard protein corona compounds allocate the biological identity of the NPs and coronas trigger NP bioactivities such as membrane adhesion, infiltration mechanism, cellular signaling pathways, biodistribution, and protection [28-31]. Not only the surface features and dimension of NPs and pH of the medium mediate their behavior *in vivo*, but also the type of proteins attached on the NP [32]. Maintaining or denaturing the native conformations of proteins after the adsorption on the NP surface could lead to the different consequences, such as effectiveness or toxicity. Thus, exploring the specific protein – NP interactions and the corresponding parameters determining the mediated conformational alterations of proteins during interactions are the pivotal features needed to explore how the biological systems could react *in vivo* [33]. Besides the protein corona formation, one aspect of the NP fate in biological environments is addressed by their colloidal stability. Recent reports suggest that NP aggregation induce a major influence on the NP-cell interactions with regard to concentration, cellular uptake and adverse effects [34]. For example, the interactions lipids NPs of different dimension and their surface modification were explored to examine the correlation between NP physico- chemical features and NP-cell interactions. It was revealed that physicochemical characteristics of NPs alone could not determine the interaction of NPs with human cervical cancer cell line (HeLa) [35]. Remarkably, determined protein corona fingerprints showed several unregulated protein adsorptions on the plasma membranes of HeLa cells [35, 36]. As shown in Figure 1, proteins can alter their structures when binding to NPs and these NP induced conformation changes are determined by the physiochemical NP surface features, and the pH of the medium.

In recent years, several studies explained dimension and surface morphology-dependent protein conformational changes induced by nanomaterials [37-41]. However, the outcomes of these studies were inconclusive. A number of bio-molecular coronas were developed in nanomedicine for drug targeting purposes. The quantity of publications relating results of investigations of protein coronas has enhanced during the last 6 years (2012–2018) considerably. The majority of papers have concentrated on the explorations of protein corona compositions mediated by NP characteristics such as dimension, morphology, and surface modification [37-42]. Likewise, Experiments show that the most important factors affecting the non-specific binding of corona protein to NPs are the shape and size particles, surface charge, hydrophobicity, and the surface softness or hardness of NPs (Table 1). As such, some outcomes could be formulated how the physicochemical characteristics of NPs influence the protein corona and trigger the conformational changes of proteins. This could fill the gap in a detailed exploration of how the NP-protein corona could switch on the apoptotic pathways of the cells. For this aim, the main goal of this review is to review the advances in NP-protein corona induction in order to correlate the physicochemical features of NPs with mediated protein corona formation, protein aggregation, and disaggregation. Afterwards, the colloidal stability of NPs in the presence of protein will be reviewed. Finally, the effect of NP-protein corona on cytotoxicity will be discussed. Methods that can be employed to study the proteins adsorbed on the NP surfaces can be categorized in 3 classes as follows [43]:

1: Methods determining the NP induced-structural changes of proteins: Fourier-transform infrared spectroscopy (FTIR), CD, nuclear magnetic resonance (NMR), X-ray crystallography, and surface-enhanced Raman spectroscopy (SERS)

2. Methods quantifying the protein coronas: Liquid chromatography–mass spectrometry (LC-MS), TEM and fluorescence spectroscopy

3. Methods identifying the protein coronas: SDS-PAGE and MS [43].

## 2. Size

It was reported that the essential blood proteins adsorb strongly on the surfaces of the NPs and the binding constant ($K_b$) and the number of binding sites ($n$), heavily depends on the NPs dimension. It was also revealed that the model proteins experience structural changes during associations with the NPs and these NPs-induced conformational changes are more obvious in case of larger NPs. Table 2 summarizes the effect of NP size on the protein corona structure. Therefore, it may be suggested that adsorption of proteins onto the larger size NPs results in larger perturbations of the protein conformations. Also, a typical Langmuir adsorption model is observed for proteins on NPs with diffrenet sizes. These results are maybe due to the larger surface curvature of smaller NPs and efficient preservation of native protein conformations and subsequent potent efficacy in comparison with larger NPs. Also, it was indicated that the diameter of the attached protein layers enhances continuously as the NP size increases [44]. AuNPs of different sizes were fabricated and their interaction mechanisms with the BSA were explored by fluorescence spectroscopy [45]. Experimental outcomes displayed that BSA molecules attached to the metallic surfaces via electrostatic binding experience marked quenching of their intrinsic fluorescence and reduction of the quantum yield of a protein is unique for each NP dimensions. Indeed, reduction of the quantum yield is higher in the case of smaller AuNPs [45].

The AuNPs-induced structural alterations of bovine heart Cyt c of different dimensions were investigated by biophysical approaches [46]. The results displayed in different conformational changes of Cyt c upon interaction with AuNPs of different dimension due to the various linkages between protein and NPs. It was reported that the AuNP surface induces the NP-mediated conformational changes of the adsorbed Cyt c. More folded structure observed when Cyt c interacted with smaller AuNPs than that adsorbed on larger AuNPs, and the activity of Cyt c attached on the smaller AuNPs was higher than that attached on larger AuNPs [46]. The electrostatic interactions were demonstrated to be mainly the crucial determining factor leading to the formation of Cyt c/16-nm AuNPs complex, whereas the hydrophobic interactions were deduced to be the leading forces in the adsorption of Cyt c on 2-4 nm AuNPs (Fig. 2).

NP dimensions and their binding to plasma proteins mediate the prolonged circulation of NPs in the bloodstream, which can be considered for therapeutic potency. An approximate magnifying of AuNP hydrodynamic dimension was reported after incubation of AuNP with different sizes in human plasma [47]. The zeta potential of the 30 nm and 50 nm were reduced significantly after incubation with plasma proteins. The plasma proteins adsorbed onto the AuNPs were separated by 2D SDS-PAGE and identified by MS which revealed a higher abundance on the 30-nm particles [47].

**3. pH**
The conformational changes of proteins upon interaction with NPs were explored in detail by various spectroscopic approaches [48]. It was well documented that proteins in the complex experiences marked structural alterations of both secondary and tertiary conformation [48]. Changes in the pH of the environment may result in more or less corona protein binding to nanomaterials, by changing the protein pattern. Since, nanomaterials with protein corona coverage, in absorbing paths, are exposed to a variety of pH conditions, it is necessary to investigate the pattern of corona protein in the biological fluids of the blood (neutral pH), intracellular fluid (pH 6.8) and lysosomes (pH 4.5-5) [49]. The outcome determined that the pH of the system substantially affected the structural changes of proteins and that fluorescence and CD studies showed that the induced denaturation was even more pronounced at basic pH [48] (Fig. 3). Table 3 summarizes the effects of medium pH on the adsorption behavior of proteins. Based on the reported data, it may be suggested that NPs which are substituted with different moieties and different bonds (chemical, electrostatic, etc.) could be applied in pH- controlled release system.

PH-sensitive NP-protein conjugates can be tuned by incorporating acid or base-cleavable bonds for application in prolonged smart drug delivery systems.

The adsorption and conformation of proteins on NPs were explored by different methods [50]. The AuNP and silica NPs size distributions revealed a greater enhance due to interactions with BSA at acidic pHs in comparison with basic pHs (3.4−7.3). These larger NPs sizes triggered reversible unfolding of BSA at acidic pH. Over the acidic pH range, the size distribution of the BSA/AuNPs or silica NPs complex remains almost unchanged, indicating the structural stability of BSA over this pH range [49, 50]. Analogously, Givens, Xu, Fiegel and Grassian [42] by applying silicon NPs with albumin coating in various pH environments (range of 2 to 8), showed that the highest absorption of albumin by silicon nanoparticles was at pH 3.7, which is close to the isoelectric point of the silicon-albumin NPs. They also reported that the structure of BSA absorbed from the original BSA is different, but by changing the pH environment from 2 to the neutral, the structure of the BSA adsorbed to the original state is reversible.

## 4. Surface chemistry

The physicochemical reaction between bio-molecules and NPs are dependent on surface features of the NPs. The effect of NPs with positive, negative charges and neutral on the structure of Cyt c was investigated [51]. CD spectroscopy showed that changing the NP charges altered markedly the conformational stability of adsorbed protein. Indeed, protein retained its native structure upon interaction with neutral moieties, but experienced large folding changes on the surfaces of charged NPs (Fig. 4) [51]. This is explained by the electrostatic interactions of charged residues and ligands. Zwitterionic NPs of variable hydrophobicity showed no induction of hard coronas at physiological serum concentrations (Fig. 5) [52]. These NPs can be implemented as potential candidates to explore nanobiological fate such as cell infiltrate and membrane damage mediated directly by chemical moieties on the NP surface.

Citrate groups with negative charges at physiological pH, have been employed to functionalize the surface of particles, in particular for Au [53-55], Ag [55, 56] and oxide NPs [57-59]. In the presence of serum protein, the citrate-modified NPs lost their colloidal stability [53, 54, 56-60]. The colloidal destabilization kinetics of citrate capped magnetic NPs in cell culture medium without serum was attributed to their complexation with counterions. Serum protein due to formation of protein corona delayed the aggregation of NPs [58, 59]. To decrease the protein corona impacts, reports have exhibited that neutral polymers or polysaccharides are efficient [61]. Tethered at the NP surface by the chain extremity, the polymers form a swollen brush that provides a protective layer against protein attachment. A number of approaches based on physical or chemical binding are now attainable, including spontaneous adsorption, layer-by-layer deposition or surface-induced polymerization. Non-covalent reactions by the assembly of separate parts (e.g. NPs, ligands, polymers) reveal increased yields based on the quantity of NPs fabricated and are prone for implementations [62]. Recent studies have exhibited that PEGylated NPs from oxides or noble metals [60, 63-67] demonstrate reduced protein adsorption. By comparing identical core NPs with different surface moieties, it was displayed that the optimum conditions for mitigating the protein corona are PEG chain with densities around 1 $nm^{-2}$ and molecular weight between 2000 and 10000 g $mol^{-1}$ [64, 67, 68]. The exact mechanisms of the protein corona on polymer brushes have not been elucidated yet. Table 4 outlines the impact of surface modification of NPs on the adsorption behavior of proteins.

It was deduced that, the surface heterogeneity of NPs allows the protein to adopt different conformations during interaction. In the case of NPs with negative charge groups, positively-charged side chains form electrostatic interactions with the negatively-charged moieties on the NP surface, while nonpolar side chain clusters may bind with the nonpolar sites on the surface of NPs. Overall, the results of this part indicated that nanoscale manipulations of NP surfaces play a vital role in controlling of the NP-protein adsorption.

## 5. Morphology

The interaction of NPs with proteins also depends on the morphology of NPs. The effect of NP morphology on the conformational changes of lysozyme and α-chymotrypsin was explored [69]. It was revealed that proteins ligands more adsorb on the surface of Au nanorods when compared with Au nanospheres. Because of raising lateral interactions on the comparatively 'flat' cylindrical surface may enable a higher packing concentration of proteins adsorbed on Au nanorods in comparison with Au nanospheres. After the interaction of lysozyme with Au nanospheres and Au nanorods, the more secondary structure was determined after interaction with Au nanorods relative to Au nanospheres, causing to complex aggregation and resulting remarkably decreased enzymatic activity. ChT retained its secondary conformation and subsequent activity during interactions with Au nanorods and Au nanospheres. Therefore, it was suggested that AuNP morphology as well as the kind of proteins influence the adsorbed protein folding [69].

Spherical AuNPs and flat Au films were fabricated with yeast Cyt C covalently attached to the Au surface. Upon exposure to different pH values, it was revealed that Cyt C denatures at acidic pH and refolds at basic pH irrespective of the morphology of NPs [70]. The binding between BSA and spherical or nanorod AuNP was characterized by different optical spectroscopy approaches. There was a superior adsorption of BSA to the two ends of the Au nanorods. The binding assay showed that the binding constant of BSA to spherical AuNPs is higher than in the case of nanorod AuNPs. The spectroscopic data exhibited the molecular specificity of complex and feasible conformational changes of BSA on the surface of AuNP (Fig. 6) [71]. This data allows better understanding of the adsorption mechanism of proteins on the surface of NPs with the different morphologies applied in pharmacy or medicine.

## 6. Protein corona induced NP aggregation

Herein, we emphasize a feature that is not always addressed in NP *in vitro* and *in vivo* investigations, namely the colloidal stability of the NPs in biological fluids. Besides the protein corona formation, several groups have reported that *in vitro* NPs aggregate and form clusters of various sizes. Recent reviews are specifically focusing on the NP fate in biological systems and on their effects on dosimetry, cellular infiltrate and induced cytotoxicity [34, 72].

### 6.1. Exchange and/or the removal of the protecting coat

NP aggregation has been determined for a broad range of NPs, including oxides (Ce, Fe, etc…) [58, 73-78], noble metals (Au, Ag) [55, 73, 77, 79, 80], semi-conductor quantum-dots [81] or polymers based NPs [82]. In these studies, the aggregate dimensions are characterized by light scattering (Fig. 7a and 7b), but their internal structure and morphology are in general not determined. Although, a precise description of NP aggregation is still unknown, it is well documented that the agglomeration of NPs in biological systems is deduced from the exchange and/or the removal of the protecting coat or from the adsorption of biological macromolecules (including proteins) at the NP surface [34, 59]. In the former case, the initial coating is not efficient to overcome the van der Waals interactions, leading to agglomeration and the sedimentation processes depend on various factors such as the nature and robustness of the coat [59]. In the latter case, the adsorbed macromolecules are playing the role of physical bridges between NPs [34]. As illustrated in Figure 7c and 7d using TEM, the NP aggregation plays an pivotal role in the NP-cell interaction, and internalization.

## 6.2. Reaction-diffusion processes

Reaction-diffusion processes are here responsible for the building of large disordered and non-colloidal structures. In the more general context, "what the cell will see" in case of NP aggregation is not inorganic cores coated or not with proteins, but rather NP clusters with proteins and other biomacromolecules at their outer layers (Fig. 8). The hydrodynamic features of such aggregates are also drastically altered with respect to single NPs, and so will be their interactions with cells. Sedimentation effects of micron size aggregates have been put under scrutiny recently, in particular using the *in vitro* sedimentation, diffusion and dosimetry model developed to improve dose metrics methodologies [83]. This model finds that the doses administered to cells *in vitro* depend predominately on the NP dimension and on their densities. Because of gravity there is a concentration gradient of NPs that forms above the cell layers. In relevant cases it was explored that the effective concentration delivered to the cells can be much higher than the nominal one by a factor 10 to 100, initiating a bias in the cellular uptake and toxicity dose dependencies. In conclusion, it seems now well established that following the protein corona formation, other phenomena such as aggregation and sedimentation occur when NPs are dispersed in physiological conditions, and that the *in vitro* and *in vivo* interactions depend heavily on the NP colloidal state. Figure 8 illustrates the basic scenarios for aggregated and dispersed NPs in interaction with cells [59, 84].

## 7. Protein aggregation

NPs can speed up dramatically the rate of protein fibrillation or the formation of fibrils inducing amyloid diseases by diverse mechanisms. It was observed that NPs initiate protein aggregation at physiological pH, causing the induction of unfolded, amorphous protein-NP complexes, followed by large protein clusters [85]. The NP-initiated conformational changes of proteins very likely catalyze the formation of aggregate species and extension [85].
It was shown that NPs such as copolymer NPs, cerium oxide NPs, quantum dots, and carbon nanotubes increase the formation of an essential nucleus for the formation of protein fibrils from human β2-microglobulin [86]. The detected shorter nucleation phase is heavily controlled by the concentration and NP chemistry. Proteins at the NP surface may experience different conformational changes and subsequent amyloid extension. It was also observed that Aβ fibril formation by $TiO_2$ NPs is promoted. Indeed, NPs due to their high accessible area could shorten lag phase as the key rate-controlling phase of fibrillation [87]. Therefore, the impact of NP on amyloid aggregation depends on a number of parameters such as protein stability, intrinsic aggregation tendency, surface chemistry of NPs, pH of medium, and type of NP.

## 7.1. Protein concentration

It was revealed that β2-microglobulin adsorbs in the form of several layers on the NP surface. Such a locally enhanced protein concentration on the NP surface can facilitate oligomer extension. This and the shortened lag phase of aggregation mediate surface-derived nucleation enhancing the risk for toxic aggregated products and amyloid formations [86]. Surface Plasmon resonance (SPR) study revealed that β2-microglobulin adsorbs on the N-isopropylacrylamide/N-tert-butylacrylamide (NIPAM/BAM) NPs in the form of several layers. Therefore, it may suggest the aggregate formation in the presence of high concentration of protein [86, 88].

## 7.2. Surface charges, NP concentration and pH

It was demonstrated that the thickness and the surface groups of the SPIONs can influence the kinetics of fibrillation of Aβ *in vitro* [89]. Concentration and surface modifications can play a dual impact on protein fibrillation. While, lower concentrations of SPIONs delayed the nucleation phase, its higher concentrations enhanced the rate of Aβ amyloid formation. With respect to surface charge, it was explored that the positively charged SPIONs can initiate fibrillation at dramatically low NP doses compared with negatively charged or uncharged SPIONs. This data indicated that in addition to the NP concentration, surface modification of NP also affects the protein conformation and corresponding formation of aggregated species. It was also reported that AuNPs initiate protein aggregation at physiological pH, causing the formation of unfolded, amorphous protein- NP complexes without embedded NPs (Fig. 9) [90].

## 7.3. Intrinsic protein stability

Indeed, proteins at the NP surface experience local destabilization inducing the formation of aggregate species and their extension. The impact of NP on amyloid aggregation heavily depends on a number of parameters such as protein stability and intrinsic aggregation tendency [91]. The aggregation process was explored in the absence and presence of copolymeric NPs with varying hydrophobic moieties. It was found that in the case of proteins with a very stable structure and low intrinsic aggregation tendency fibril formation is promoted by NPs. However, in case of low stable proteins and high intrinsic aggregation tendency, an inhibitory impact of amyloid fibril induction by NPs was determined (Fig. 10) [91]. Fluorinated NP was demonstrated to induce no Aβ40 aggregation, whereas hydrogen substituted NPs promoted the conversion of Aβ40 from natively unfolded structure to β-sheet clusters [92]. Therefore, a valid association between the protein stability and the NP impact on the amyloid formation can be determined. It was shown that the native state of monellin protected it against aggregation [93].

## 8. Amyloid disaggregation

NP with substituted functional groups can be applied as potent ligands in preventing and curing of protein aggregation-derived diseases. The nucleation phase (rate-controlling step) is heavily dependent on surface modification of NPs. Generally, nanomaterials have very high surface energy and a strong tendency to each other [94]. Especially for those who have OH levels on their surface [95]. Therefore, by changing the functional levels of the nanomaterials by a variety of compounds, their agglomeration can be prevented. As a result, with the reduction of the growth phase, the nucleation stage increases [96].

## 8.1. Surface chemistry

Amyloid clusters of Aβ42 were dissolved via the combined application of microwave irradiations and AuNPs. AuNPs with H-Cys-LeuPro-Phe-Phe-AspNH$_2$ (Cys-PEP) as an attached peptide selectively binds to the Aβ aggregates, inducing the conjugated Au-NP-Cys-PEP. This peptide can recognize hydrophobic patches in the amyloid structure [97]. Polymer NPs with different hydrophobicity was revealed to delay fibrillation of Aβ [98]. It was indicated that these NPs influence primarily the nucleation phase of Aβ aggregation. The extension phase is not normally influenced by the NP. However, the extension phase was reported to be markedly dependent on both the dose and surface characteristics of the NPs. SPR outcomes displayed that the adsorption of monomeric Aβ onto the NPs inhibits fibrillation. Moreover, it was revealed that fibrillation of Aβ can be reversed in the presence of NPs [98].

The adsorption of protein on the NP surface decreases the concentration of free protein in solution to initiate the aggregation process, extending the lag phase of aggregation. Additionally, the reduction of aggregation was observed by the equilibrium phase at the end of the fibrillation pathway (Fig. 11) [99]. LVFFARK (LK7) in the absence and presence of poly (lactic-co-glycolic acid) (PLGA) were designed to explore the amyloid β-protein fibrillation. LK7 was detected to inhibit Aβ42 fibrillogenesis in a concentration-dependent system. However, LK7@PLGA-NPs demonstrated dramatic inhibitory effect against Aβ42 aggregation [100]. It was also revealed that the fibril fibrillation can be retarded by inhibiting the nucleation and extension phases with a low concentration of functionalized quantum dots. The inhibiting impact of functionalized quantum dots on the fibril formation revealed that the functionalized quantum dots concentration was low [101].

Dendrimers with anti-aggregation features are regulated both by surface chemistry and surface group density. Dendrimers with cationic groups are more potent in inhibiting amyloid β-protein aggregation than surface hydroxyl groups and unmodified dendrimers [102]. A dopamine-containing PEG-based polymer with high density was developed for the retardation of α-synuclein fibrillation [103]. Polymeric NPs can inhibit Aβ fibrillogenesis depending on their surface characteristics [104]. It was found that polymeric NPs inhibit the amyloid extension due to negative surface charges. It was deduced that negative charge distribution van strength the stretching of Aβ42 molecules instead of the formation of amyloid aggregates [104].

Green tea polyphenol epigallocatechin-3-gallate (EGCG) in the form of EGCG NPs (nano-EGCG) are more influential than molecular EGCG in retarding amyloid formation, destabilizing mature protein amyloids, and reducing amyloidogenic cytotoxicity (Figure 12) [105]. It was reported that sugar-modified AuNPs inhibit intracellular polyglutamine-containing mutant protein amyloid elongation (Fig. 13) [106]. Sugar-terminated NP chaperones were observed to be more dynamic than molecular sugars in retarding protein fibrillation and reduced adverse effects (Fig. 14) [107]. It was demonstrated that fibrillation of HSA is rerated by adsorption on the chitosan- modified AgNPs [108]. Synergistic impacts of negatively charged hydrophobic NPs and (−)-epigallocatechin-3-gallate was reported on inhibiting of Aβ elongation [109]. It was proposed that iminodiacetic acid-conjugated NPs (IDA-NP) can control Aβ42 amyloid formation and mitigate the cytotoxicity induced by $Zn^{2+}$ [110]. Interaction of proteins with NP with high local protein density can play a pivotal role in amyloid formation. On the other side, protein adsorption can modulate the concentration of free protein in solution and consequently results in an extended lag-time for protein cluster formation. Indeed, interaction of NPs and proteins could block the active sites for fibril induction and can mitigate free protein population in solution, thus inhibiting formation of fibrils. Certainly much more data are needed to define or foresee the effects of NPs on fibrillogenesis.

## 9. NPs-protein corona and cytotoxicity

Although NPs have immense application in industry, scientific knowledge and biomedical research, it can influence the biological nature of the cellular and subcellular levels. Distinct "protein corona signature" is formed by different protein component and is associated with the dimension and/or surface functional groups of the NPs. So, the induction of NP-protein coronas together with the biological reaction to these cellular biophysico-chemical mechanisms (i.e, biotransformation, biodistribution and endocytosis,) and coronas (i.e., cytotoxicity, oxidative stress and inflammation) will be crucial for nanomedicine and nanotoxicology. The cytotoxicity of NPs is heavily modulated by their interactions with blood proteins.

### 9.1. NPs interactions with plasma factors and proteins
NPs interactions with plasma factors and proteins can influence the clearance and systemic adverse effects of the NP delivery system [111]. The adverse effects deriving from blood interactions are often ignored in the design and fabrication of NP-based systems for *in vivo* implementations. These toxicological threats are due to their small size as it can penetrate cell membranes and cross the various biological barriers and may reach the most sensitive organs [112]. Hazardous effects include protein denaturation induced by the large accessible area of NPs and DNA breaks in human cells induced by carbon nanotubes [113]. The NP-bound blood proteins can affect the immune system by stimulating immune cell reactions. Moreover, protein corona considerably influences interactions between NPs and cells through different activation of pathways along with internalization. For this purpose, in order to study the level of toxicity of NPs and its effects on cells, computational methods are used in addition to experimental methods. The powerful computing software such as the quantitative structure–activity relation (QSAR) is emerging. In these applications, there is a basic condition that most nanomaterials with comparable physical and chemical properties (size, zeta potential, particle surface area, moles of particles, and protein density) have almost the same biological activity. For example, Gajewicz, Schaeublin, Rasulev, Hussain, Leszczynska, Puzyn and Leszczynski [114] developed a nano-QSAR model for the toxicity of 18 kinds of metal oxide NPs for keratinocyte (HaCaT) cell line, which is a common model for facing NPs with skin. Also, Manganelli, Leone, Toropov, Toropova and Benfenati [115] designed a QSAR model that accurately predicts the human embryonic kidney cells (HEK293) response in the presence of 20 to 50 nm silica NPs. In another study based on the QSAR model and the 7 types of oxide NPs, it was shown that increasing the measured properties of NPs, especially the nanoparticle network model, the dosage, the enthalpy of formation, the exposure time and the hydrodynamic size, could increase the prediction level of the model by more than 90% [116].

The greater surface area of NPs results in their high chemical activity and subsequent enhanced reactive oxygen species (ROS) production causing oxidative stress, inflammation [117, 118]. The mechanism of NP-induced cytotoxicity caused by non-soluble metal oxide NPs are still not clear, but reports suggest that these NPs stimulate oxidative stress, genotoxicity and immunotoxicity [119]. Gold NPs are used for cancer diagnosis and treatment [120], but it has been seen that gold NPs exert toxicity through the induction of oxidative stress and mitochondrial damage [121]. From all this, we inferred that the formation of NP-protein coronas significantly influence the immunological response by the physiochemical surface features of the NP and it should be addressed in future works by investigating the protein distribution and exploring *in vitro* and *in vivo* responses.

### 9.2. The impact of NPs-protein corona on cytotoxicity
As mentioned, the NPs via a disconnection of the totality of plasma membrane, mitochondrial collapse and nucleus injury may cause cell death (Fig. 15) [122], which can be greatly reduced by the physicochemical features of the protein corona like inaccessibility to the surface, the modification in the surface charge of NPs, etc. and its progression from soft to hard condition with low dynamic exchange in biological conditions [123]. It is clear from the data of figure 15 that the protein corona significantly reduces the absorption of NPs by cells. While corona protein surface modification can reduce the absorption of nanomaterials. In particular, absorption suppression is visible for HSAsuc-coated NPs. While the presence of ethylenediamine in the protein corona can increase the cell wall uptake compared with the HSA native. However, intracellular penetration is also significantly reduced in comparison with the native group, which may be due to electrostatic interactions between NPs and cell surface. Indeed, with the creation of corona protein on NPs, the immune response to the NPs has altered and NPs become more biocompatible to the human body, whereas the protein corona increases the targeting and delivery of NPs in some cases. Since the toxicity of NPs generally arises from the contact with the membrane cells [124], creating a coating such as corona protein can help eliminate this problem and increments the NPs safety. In this respect, some research has shown that the toxicity of graphene oxide nano-sheets by FBS and BSA has been intently decreased due to the prevention of interface between NPs and cells (Fig. 16) [125, 126].

Their outcomes displayed that protein corona decrements the cytotoxicity through a remarkable careless of the graphene and membrane lipid contact and a decrease of the accessibility level area of graphene oxide. These findings provide different visions into the physical interaction and potential offend between NPs and cell membranes. Also, experiments showed that cells exposed to silica NPs without corona protein (HSA) had a developed degree of adherence to the cell membrane and larger permeability [127]. Besides, Yin, Chen, Casey, Ke, Davis and Chen [128] exhibited that the covering of zinc oxide NPs with corona protein decreases their cytotoxicity into skin fibroblasts and hepato-carcinoma human. Another study showed that the usage of corona protein as a covering of silver NPs reduced cytotoxicity by affecting cell surface receptors, whereas increasing the intracellular uptake of silver NPs [129]. Along with, it was shown that silver NPs prepared with carbohydrate covering shared with corona proteins created very low cytotoxicity in neuron-like cell line and a hepatocyte cell line compared to carbohydrate-based silver NPs [130]. Despite the high toxicity of carbon nanotubes, Ge, Du, Zhao, Wang, Liu, Li, Yang, Zhou, Zhao and Chai [131] and De Paoli, Diduch, Tegegn, Orecna, Strader, Karnaukhova, Bonevich, Holada and Simak [132] confirmed that the binding of blood proteins like albumin, histone, etc. to CNTs has greatly reduced their cytotoxic effects on umbilical vein endothelial cells and platelets. Although, reducing the NPs cytotoxicity due to decreased cellular connection and uptake by corona protein have been reported, this reduction can be considered as a negative point when using NPs for drug transfer into cells [125]. For instance, Ritz, Schöttler, Kotman, Baier, Musyanovych, Kuharev, Landfester, Schild, Jahn and Tenzer [133] revealed that NPs enveloped with the apo-lipoproteins ApoA4 or ApoC3 significantly declined the cellular uptake into human mesenchymal stem cells, while in the same experiment, it was shown that the covering of NPs with ApoH increased the cellular uptake. Accordingly, the superficial adjustment process by corona proteins requires to be intently noticed and optimized realize for the greatest therapeutic effects.

The interface between NPs and corona protein might decrease systemic toxicity. This scenario, similar to previous toxicity, is seriously reliant on the inherent level of NPs such as the creation of ROS owing to surface chemical activity [134]. The corona protein can augment the safety of NPs predominant materials with semiconductor properties by prohibiting the output of ROS. On this basis, Tenzer, Docter, Kuharev, Musyanovych, Fetz, Hecht, Schlenk, Fischer, Kiouptsi, Reinhardt, Landfester, Schild, Maskos, Knauer and Stauber [74] declared that uncovered silica NPs had a substantial hemolytic effect on red blood cells, whereas the blood cells were protected from destruction by the NPs due to the presence of corona protein. In addition, zinc oxide NPs are an obvious fabricator of ROS due to their pristine surface, but the presence of a corona protein prevents this occurrence [128]. Also, *in vitro* studies offered that AuNPs performance as cytotoxic factors are related to an increment in ROS in human liver hepatocellular carcinoma cells, whereas AuNPs treated with corona proteins derived lung fluid were not toxic to human hepatoma cell line [135]. In concurring with this experiment, [22] showed that the use of corona protein like HSA and human plasma protein as a coating of AuNPs can moderate the cytotoxic effects of ROS and uptake cellular of NPs on hepatocyte cells. Nevertheless, corona proteins occasionally increase the cytotoxicity of a NPs due to protein denaturation with alterations of constancy and structural. The greatest important examples in this field are the contact of AuNPs with fibrinogen proteins as corona protein in the blood causes inflammatory responses when they are in interface with the leukocyte receptor MAC-1 [136]. As well as, the unwanted changes of tubulin protein as corona protein in the presence of titanium dioxide NPs is another example that reduces the ability of tubulin polymerization [137]. In conclusion, exploring and realizing the adverse effect of nanostructure-based substances treated with protein coronas is a serious assay toward controlling their cell uptake and subsequent oxidative stress and toxic outcomes (Fig. 17).

## 10. Clinical application of protein corona-NP

As mentioned, NPs have a low stability and high systemic toxicity or cytotoxicity to the cells due to its high free surface energy. Therefore, biomolecules such as corona protein are used to increase the stability and decrease the toxicity or cytotoxicity of NPs. In addition, based on the information obtained from papers, corona protein can be very effective in targeting NPs for drug delivery and imaging.

## 10.1. The impact of NPs-protein corona on drug delivery

One of the most important questions in the field of NPs cellular uptake is to increase their uptake with specific targets. Nonspecific uptake has been always a random process, whereas the specific uptake of NPs into a cell depends on cell surface membrane receptors and functional agents on the NPs. In this regard, several studies have shown that corona protein is highly effective in the targeted uptake of NPs and the stability of NPs due to preventing from renal filtration [138], but the key challenges of NPs in the body are the presence of corona protein compounds in the blood plasma, which can change the nature of the activity of NPs [139] and cause lag drug release by prohibiting drug diffusion via the matrix [140] along with decreasing of the burst effect [141]. Whereas corona protein can increase the loading capacity of a drug like a sponge as well as control the release time of the drug by simultaneously loading multiple drugs, and photothermal and photodynamic substance for cancer treatment [142]. Accordingly Qi, Yao, He, Yu and Huang [143] developed dextran–chitosan NPs coated with BSA along with doxorubicin that in addition to reducing the cytotoxicity, increasing the anti-tumor effect of doxorubicin and survivability of mice. However, Kah, Chen, Zubieta and Hamad-Schifferli [144] showed that protein corona despite its positive effect on the release of doxorubicin drug in the target cell, may have undesirable effects on gene-dependent therapies due to reduced release of genetic compounds in cells. In contrast, Kummitha, Malamas and Lu [145] showed that the corona proteins application on siRNA lipid NPs results in the improvement of the siRNA delivery in the body due to prevent non-specific association of serum molecules. In another study displayed that the use of lipid NPs coated with HSA in gene delivery to treat breast cancer cells was much more effective than control groups [146]. Generally, Cifuentes-Rius, de Puig, Kah, Borros and Hamad-Schifferli [147] demonstrated that the release rate of genetic compounds such as DNA loaded on the NPs could be modified by changing the type of corona protein. Analogously, Caracciolo, Cardarelli, Pozzi, Salomone, Maccari, Bardi, Capriotti, Cavaliere, Papi and Laganà [148] revealed that the use of vitronectin-integrin as a corona protein on lipoplexes NPs made of 1,2-dioleoyl-3-trimethylammonium propane and DNA, increases the cellular uptake of nanoparticles in HEK 293 and MDA-MB-435s cell lines by up to twice as much as the control group. They also showed that corona protein increase the survival of nanoparticles in the blood and reduce toxicity. In this line, Digiacomo, Cardarelli, Pozzi, Palchetti, Digman, Gratton, Capriotti, Mahmoudi and Caracciolo [149], in a cellular model, were able to cellular uptake of the provided NPs with the higher performance in HeLa cell lines, by developing liposome NPs coated with apolipoprotein as protein corona derived from human plasma. It has been shown in this study that apolipoprotein as a protein corona induces NPs into target cells via macropinocytosis to clathrin-dependent endocytosis.

In the following, Schäffler, Sousa, Wenk, Sitia, Hirn, Schleh, Haberl, Violatto, Canovi and Andreozzi [150] declared that the AuNPs coated with albumin represented more accumulation in the lungs, spleen and brain in comparison with the control AuNP. Using this finding, it is possible to overcome the pulmonary diseases like tuberculosis, or to prevent bloodshed and brain diseases by crossing the blood-brain barrier. Without the use of targeting agents that corona protein pattern affects them, the application of corona protein decreases the effect of the NPs containing cancer drug [151]. For instance, Yallapu, Chauhan, Othman, Khalilzad-Sharghi, Ebeling, Khan, Jaggi and Chauhan [152] showed that the change in the corona protein from HSA to Apolipoprotein-E in the NPs can carrier the drug towards the kidney and the liver, despite the presence of cancer receptors (Fig. 18).

On the other hands, Rastogi, Kora and Arunachalam [153] demonstrated that gold NPs coated with BSA corona protein are very suitable carriers for transferring antibiotics like gentamicin, neomycin, etc. to the target tissue to eliminate gram positive and gram-negative bacteria compared to control groups. NPs containing bio-corona proteins may cause the inflammasome in macrophages, which will motivate the production pro-inflammatory cytokines cascade. Hence, they can be effective in providing vaccines. In this field, the use of coronavirus antigens in birds as corona proteins has improved vaccination compared to common vaccination [154]. Inoculation the AuNPs coated with virus-like particles as corona protein caused the elevated lymphatic antigen transfer, enhanced cell response, and decreased infection related signs in an avian model with coronavirus infection compared to insemination with free proteins. In the next study, Chang, Diambou, Kim, Wang and Champion [155] indicated that ovalbumin NPs coated with adjuvants-flagellin and immunoglobulin corona proteins derived from salmonella (disease agent) and mice (host representative), respectively, increased the cells immune response in mice.

**10.2. The impact of NPs-protein corona on cancer therapy**

For the treatment of cancer, which is one of the most lethal diseases, the chemotherapy is still the most widely used treatment method despite the many disadvantages such as high toxicity to non-target cells, non-specificity, and so on. Accordingly, nanomedicine by introducing NPs coated with drugs and biological compounds such as corona proteins has been used to treat the cancers more appropriately, with reduced cytotoxicity and increased drug targeting in clinical and laboratory levels. In this field, Caracciolo [156] explained that the use of NPs coated with protein corona such as protein coated liposomes could be a desirable option for chemotherapy activities in transferring drug to target cells without causing toxicity in other tissues and cytotoxicity due to accumulation of inorganic NPs. For example, Doxil® was originally approved for Kaposi's sarcoma (1995) and later for breast cancer (1999) and ovarian cancer (2003). However, chemotherapy drugs are usually limited within the liposome. As a result, they play a minor role in tumor cells. Therefore, attention to metallic NPs, despite specific toxicities, increased to synchronize therapeutic activities such as chemotherapy, radiotherapy, and imaging. For instance, Chunfu, Jinquan, Duanzhi, Yongxian, Yanlin and Jiajü [157] using HSA corona protein as a coating on magnetic NPs, in addition to reducing the side effects, increased the $^{188}$Re radioisotop loading capacity in cancer cells for cancer treatment by radiotherapy. Furthermore, Quan, Xie, Gao, Yang, Zhang, Liu, Lin, Wang, Eden and Lee [158] expanded HSA-coated iron oxide NPs with anti-cancer drug (doxorubicin) which corona protein due to increase of stability and higher targeting contributed to the further translocation of doxorubicin across the breast cancer cell membrane to its accumulation in the nucleus. Meanwhile, Xing, Bhirde, Wang, Sun, Liu, Hou and Chen [159] explained that hollow iron oxide NPs coated with HSA had the greatest potential for transferring doxorubicin drug to cancer cells of resistance to treatment compared to free NPs, and can simultaneously transport multiple drugs to the target cell and release them during treatment slowly along with greater power and clarity of target tissue imaging. In this line, Hajipour, Akhavan, Meidanchi, Laurent and Mahmoudi [160] showed that Superparamagnetic zinc ferrite spinel-graphene nanostructures coated with HP corona protein can have a significant effect on the treatment of cancer cells and the toxicity associated with the use of NPs by inducing the production of ROS in the target cell. Also, Arvizo, Giri, Moyano, Miranda, Madden, McCormick, Bhattacharya, Rotello, Kocher and Mukherjee [161] illustrated that with the use of corona protein along with hepatoma-derived growth factor on AuNPs, the removal of ovarian cancer cell lines has been remarkably increased. Moreover, by using albumin as a coating for lipid NPs containing Docetaxel, it was provided more transfer of the drug from the blood-brain barrier, increase apoptosis in cancerous cells, and drugs accumulate at the experimental glioma site [162]. Furthermore, Akrami, Khoobi, Khalilvand-Sedagheh, Haririan, Bahador, Faramarzi, Rezaei, Javar, Salehi and Ardestani [163] developed a magnetic iron oxide NPs with an anticancer drug as curcumin coated with HP corona protein that reduced the growth and proliferation of breast cancer cells compared to free NPs. Since the NPs less than 5 nm exhibit the lowest toxicity due to fast filtration of the kidneys, AL-Jawad, Taha, Al-Halbosiy and AL-Barram [164] produced small AuNPs (<5 nm) coated with BSA to increase stability (with increasing particle size) which in addition to increasing photothermal therapy by laser irradiation in cancer cell lines, showed the least toxicity (up to 70%). Similarly, Hai, Piraux, Mazario, Volatron, Ha-Duong, Decorse, Lomas, Verbeke, Ammar and Wilhelm [165] exhibited that the maghemite NPs coated with HSA corona protein eliminated cancer cell lines by photothermal therapy with very low toxicity. Moreover, Azizi, Ghourchian, Yazdian, Dashtestani and AlizadehZeinabad [166] showed that the CuNPs coated with BSA corona protein had the lowest toxicity and the highest inhibitory power to breast cancer cells up to 90%. Likewise, Azizi, Ghourchian, Yazdian, Bagherifam, Bekhradnia and Nyström [167] with the use of albumin corona protein on the silver NPs was able to inhibit the growth of breast cancer cells by producing more ROS, delaying release of the drug, and the stability of NPs compared with free NPs. In addition, they showed that the of silver NPs coated with HSA accumulated in the cancer cells, despite being less toxicity than the control group [167]. In line with the medication, Yeo, Joshua, Cheah, Neo, Goh, Kanchanawong, Soo, Thong and Kah [168] reported that the use of corona protein for simultaneous application of doxorubicin drug, photodynamic and photothermal agents has increased drug uptake and the removal of oral cells carcinoma by up to 98.7% with the least cytotoxicity due to minimal drug intake (Fig. 19). In other studies showed that the application of corona protein like fetal bovaine serum or HSA on AuNPs and iron NPs for simultaneous loading of the cancer drug and photothermal materials could reduce the dose of the drug, despite its an increase in the tumor after 6 hours of injection [169, 170]. However, despite the presence of gold nanorods containing drugs in other tissues such as the liver and spleen, it did not have toxic effects during the trial period (39 days). In the following, the application of hyaluronic acid and BSA corona proteins on the surface of super paramagnetic iron oxide NPs in addition to increasing the stability of NPs resulted in the appropriate transfer of paclitaxel to cancer cells compared to the free NPs [171]. In general, due to the conformity of the nature of corona protein to the body's biological conditions, the application of these biological components can increase the amount of NPs present in the body and improve blood circulation time. For instance, a 6 times increment of half-time was introduced by Peng for BSA-coated NPs compared to free NPs [140]. In this line, Guo, Sun and Zhang [172] described that lipid coated AuNPs, after being injected into the blood and combining with the corona protein derived from blood plasma, were more likely to accumulate and survive in the liver cancer cells.

## 10.3. The impact of NPs-protein corona on imaging

The imaging by NPs is one of the most important methods used in the field of nano-medicine to reduce the error rate in detecting the location of a complication. But, in the first stage, uniform distribution in the body without systemic toxicity or cytotoxicity, in the second phase, focusing in the target tissue with sustainability, and in the final stage of imaging clarity are NPs challenges which are presumed to overcome by the creation of corona protein on NPs. In this regard, the negative effect of corona proteins like HSA on the nature of NPs imaging was initially investigated and showed had no negative effect on MRI imaging, whereas increasing the capacity of the imaging labels. Accordingly, Kresse, Wagner, Pfefferer, Lawaczeck, Elste and Semmler [173] and Högemann-Savellano, Bos, Blondet, Sato, Abe, Josephson, Weissleder, Gaudet, Sgroi and Peters [174] declared that the use of transferrin as a corona protein along with thioether on para-magnetic iron oxide NPs increases the accumulation of these particles in breast cancer cells due to the presence of more receptors of transferrin on them. Despite showing promising contrast-enhancing effects transferring compounds are removed rapidly by the reticuloendothelial system. Hence, the use of these compounds does not appear to be desirable in imaging. While, the use of on para-magnetic iron oxide NPs with HSA protein coating has increased the nuclear magnetic resonance imaging of the liver with a very low dose [175]. Then it was shown in the next researches that with the use of HSA corona protein, the stability of a magnetic NPs containing the imaging label has increased by up to 90% (up to 72 hours), and due to the high concentration of NPs-labeled in the target tissue a higher resolution of imaging was obtained (Fig. 20) [157]. Moreover, Huang, Xie, Chen, Bu, Lee, Cheng, Li and Chen [176] exhibited that the use of HSA corona protein on MnO NPs along with $^{64}$Cu radioisotope leads to more prominent T1 contrast, 5 times transverse reflaxivity r2 and an increase in imaging. Similar to this study, Xie, Chen, Huang, Lee, Wang, Gao, Li and Chen [177] showed that the use of albumin coverage along with dopamine and $^{64}$Cu labels for imaging with iron oxide NPs increased transparency in imaging and more drug delivery to target cells with very low toxicity (Figure 18). Huang, Wang, Lin, Wang, Yang, Kuang, Qian and Mao [178] also explained that the iron NPs coated with corona protein derived from milk (casein) leading to much greater biocompatibility, stability and more accurate imaging of the breast cancer tissue. Casein protein increased the MRI contrast by enhancing the ability of T2 and developing the 2.5-fold transverse reflaxivity r2 compared to free NPs in 3 tesla. In addition, Nigam, Waghmode, Louis, Wangnoo, Chavan and Sarkar [179] developed graphene quantum dot with HSA particles and pancreatic cancer drugs that enhanced the imaging up to %14 and accumulation of anticancer drugs (gemcitabine) in tumors resistant to treatment.

As mentioned in the above studies, serum proteins as corona proteins are very useful for drugs delivery and imaging because of the increased stability of NPs and the possibility of targeting them. However, due to the extremely difficult and limited control of corona protein in bio-condition [174], their harmfulness in most targeting [152] and inaccessibility of the cell in the most cases [139], it leads to reticuloendothelial agglomeration that can cause unpredictable activity. On the other hand, the pathological alteration in diverse diseases can cause the formation of different protein corona in bio-condition. Thus, the notion of a "personalized protein corona" will be a contributing agent for clinical applications that are still ambiguous [160].

## 10.4. The impact of NPs-protein corona on detection

Non-invasive diagnosis and screening of diseases is a growing need in the medical field. In this regard, the use of nanoparticles can, in addition to increasing the accuracy of detection, reduce the speed and cost of detection. However, proteins in the blood plasma as corona proteins can be effective on the diagnostic activities via changing the physico-chemical properties of nanoparticles [180]. For example, it has been shown that AuNPs (~100 nm) containing citrate ligands with human immunoglobulin G (IgG) as a protein corona coating can improve the sensitivity of prostate cancer diagnosis by up to 50% [181]. Increasing human IgG levels as protein corona in the blood is associated to autoantibodies produced from prostate cancer, which is part of immune defense against the tumor. Also, Caputo, Papi, Coppola, Palchetti, Digiacomo, Caracciolo and Pozzi [182] found that human plasma coating on liposome NPs (~135 nm) in pancreatic cancer patients had a higher accumulation of protein corona than healthy people. They showed that the accuracy of detection of pancreatic cancer through the protein corona is improved by up to 90% versus the control group. Therefore, the interaction of protein corona in the blood plasma with NPs creates unique patterns based on human characteristics [183]. Patterns of protein corona produced by a cancerous patient are different from healthy subjects, which can be used to diagnose cancer at the early stages of tumor formation.

## 11. Opportunities and challenges

Exposure of the human body to NP can happen through several routes such as: inhalation, ingestion, injection or epidermal contact. Protection of NPs is vital, because in some cases even lower concentrations NPs initiate adverse effects on the biological system [184-186]. Corona formation results in mitigating the cytotoxicity of NPs, because the bare NP surface is active until the NP are cleared through phagocytosis [19, 32].

It is now well established that once in contact with biological systems nanomaterials acquire new biological features due to protein corona. The protein corona structure may differ depending on the physicochemical properties of NPs, including sizes, surface charges, morphologies, and aggregation states. Therefore, the different NPs incubated with plasma proteins from protein coronas with varying compositions giving rise to different protein corona structures and biological interactions, and outcomes. These changes, in turn, impact the biological responses of NPs, including their pharmacokinetics, biodistributions, and therapeutic results. Herein, we reviewed this concept along with recent advances in this domain, with a particular attention focused on conformational changes of adsorbed proteins.

On the other side, the protein corona plays an essential biological role by coating the NPs and masking the accessible sites of the nanomaterials. Therefore, it is important to determine the relationship between chemical functionalities of NPs and their biological activities, for example, in NP-induced targeted drug delivery systems and medicinal applications [187]. Additionally, NPs can destabilize the structure of adsorbed proteins and vice versa, and subsequently change the function of proteins and colloidal stability of NPs [188].

Exploring the basic information regarding the protein corona structure, NP-induced protein aggregation and disaggregation is crucial since the form and character of the nano-bio interphase defines the fate of NPs, what impacts the biological responses of the protein -NP complexes.

## 12. Conclusion

As indicated in Section 6, the inevitable coating of nanomaterials by protein corona after injection should be accepted and exploited. However, one of the main challenges for the application of nanomaterials within the field of nanomedicine is the complications deriving from the protein corona adverse properties. To this end, based on the topics of Sections 2 to 5, we need to have a profound understanding of the formation of protein corona and its effects on biological interactions based on the chemical composition, size, shape and surface properties of nanomaterials. Depending on the protein corona structural changes induced by nanomaterials surveyed in this perspective, one can deduce that there is a high necessity for a systematic remedy for protein corona induced complications. But it should be kept in mind that the correct binding of the protein groups can have favorable biological responses for medical applications such as target delivery and uptake, larger circulation time, activation of immune responses, evacuation of pathologic biomolecules and more importantly, reducing cytotoxic effects of nanomaterials. However, precautions should be taken to ensure that the functional levels of the nanomaterials are not filled up to prevent interactions between cell receptors and nanomaterials. Such advances in the field will not only minimize the contradictions in the nanotoxicology perception, but it could also bridge the gap between *in vitro* and *in vivo* outcomes, what could accelerate the application of nanomedicine techniques in clinics when assuring their safety.

**Author contributions**
The authors contributed equally to this work.

**Competing interests**
The authors declare no competing financial interests.

**Table 1.** The cause of non-specific binding of corona protein to nanoparticles [189].

|   | NP parameter | Observed effect |
|---|---|---|
| 1 | NP size | Bigger size, larger degree of protein coverage |
|   |   | Smaller size increases corona thickness and decreases conformational change |
|   |   | Evolution of composition and relative abundance of adsorbed proteins |
| 2 | NP shape | Higher protein adsorption onto nanorods compared to nanospheres |
|   | NP surface Charge | Charge affects composition of formed protein corona |
|   |   | More highly charged surfaces increase protein conformational changes |
|   |   | Protein conformation: positive > negative > neutral |
|   |   | Charged particles have higher cell internalization and faster opsonization rates than electrically neutral particles; positively charged NPs are incorporated by cells in higher numbers and faster than negatively charged NPs |
| 3 | Hydrophobicity | Increase adsorbed protein quantity and qualitatively change obtained protein adsorption patterns |
|   |   | Increase affinities of biomolecules |
|   |   | Increase protein conformational changes |
|   |   | Are opsonized more quickly than hydrophilic NPs |
| 4 | Smoothness/roughness | Surface roughness greatly minimizes repulsive interaction and influences the amount of protein, but not the protein identity |

**Table 2.** The impact of NP size on the behavior of proteins upon adsorption onto the NP surface.

| NP | Size (nm) | Technique | Protein | Outcomes | Refs |
|---|---|---|---|---|---|
| SiO$_2$ | 4, 20 and 100 | CD and UV-Vis | Lysozyme | Denaturation of adsorbed lysozyme | [190] |
| N-isopropylacrylamide (NIPAM): N-tert-butylacrylamide (BAM) copolymer | 70 and 100 | ITC | HSA | The equilibrium association constant was not correlated with the dimension and composition of NPs. | [191] |
| SiO$_2$ | 11, 36 and 150 | Centrifugation assays and CD | Subtilisin Carlsberg (SC) | A typical Langmuir adsorption model as well as conformation changes. | [192] |
| SiO$_2$ | 4, 15, 35 | CD | Cyt C | Little change of structure of cyt C, albeit with slightly larger changes for cyt C on larger NPs. | [193] |
| SiO$_2$ | 20, 30, and 100 | Label-free liquid chromatography MS | Plasma proteins | Absorption of 37 % of proteins | [194] |
| Negatively charged lipid-based NPs (liposomes) | 27 and 103 | NMR | Ubiquitin | Lifetime ($\tau_{ex}$) < molecular tumbling times ($\tau_R$) for 103 nm liposomes and >$\tau_R$ for 27 nm liposomes | [195] |

| NP | pH | Technique | Protein | Outcomes | Refs |
|---|---|---|---|---|---|
| SiO$_2$ NPs | 6,9,15 | NMR and CD | Human carbonic anhydrase | Larger NPs mediated larger denaturation of the protein secondary structure | [196] |

**Table 3.** The influence of pH of medium on the adsorption behavior of proteins.

| NP | pH | Technique | Protein | Outcomes | Refs |
|---|---|---|---|---|---|
| SiO2 | 2-12 | X-ray scattering (SAXS) | Lysozyme | Protein binding at different pH is irrespective of the ionic strength | [197] |
| SiO2 | 2–11 | UV-Vis | Lysozyme and ß-lactoglobulin | Repulsive protein-surface and protein–protein interactions. | [198] |
| SiO2 | 5-9 | Small-angle neutron scattering (SANS) and UV–vis | Lysozyme | Colloidal NPs existing with increasing pH | [199] |
| SiO2 | 5-11 | DLS and SANS | Lysozyme and BSA | The colloidal stability of NPs in the presence of lysozyme and NP agglomeration in the case of BSA when pH approaches its respective IEPs. | [49] |
| Micelles encapsulating BSA and streptavidin | 2-12 | UV-Vis | BSA | NP released the proteins in neutral and basic pH | [200] |
| Fe$_3$O$_4$ NPs-capped mesoporous silica | 2-12 | UV-Vis | Bone morphogenetic protein 2 | pH-responsive controlled release system | [201] |

**Table 4.** The impact of surface functionalization of NPs on the adsorption behavior of proteins.

| NP | groups | Technique | Protein | Outcomes | Refs |
|---|---|---|---|---|---|
| Magnetic NPs (MNPs) | Positive polyethyleneimine (PEI-MNPs) and (b) negative poly(acrylic acid) (PAA-MNPs) | DLS | Cell culture medium proteins | Production of large MNP-protein aggregates | [202] |
| Maghemite (Fe$_2$O$_3$) NPs | Citrate ions, poly(acrylic acid) | DLS, TEM | Serum proteins | NP aggregation initiated by protein corona formation | [58, 59] |
| Maghemite (Fe$_2$O$_3$) NPs | Phosphonic acid PEG copolymers | TEM, DLS, Quartz crystal microbalance (QCM) relaxometry, | Serum proteins | Particle devoid of observable protein corona | [64, 67, 68] |

| NP | Surface modification | Method | Protein | Outcome | Ref |
|---|---|---|---|---|---|
| Enshrouding polymer-coated NPs | PEG | FCS | HSA and fibrinogen | Adsorbed HSA molecules contrary to fibrinogen are buried | [203] |
| Fe–Pt NPs | Carboxyl, amino and both | FCS | HSA | No effect of surface modification on protein corona formation | [204] |
| CdSe/ZnS quantum dots | Carboxyl, amino and both | FCS | HSA | No effect of surface modification on protein corona formation | [204] |
| Au and Ag nanoclusters | Carboxyl, amino and both | FCS | HSA | No effect of surface modification on protein corona formation | [204] |
| Iron oxide NPs (SPIONs) | Surface hydrophobicity | Flow field-flow fractionation and ultracentrifugation | Human serum | Increasing surface hydrophobicity results in formation of soft corona | [205] |
| AuNPs | Polar and nonpolar groups | ITC | BSA | Adsorption of BSA on AuNPs was primarily caused by electrostatic interactions. | [52] |
| AuNPs | Citrate-coated | NMR | GB3 and ubiquitin | Charge, $K_b$, and dimension of each protein determine the stoichiometry of protein adsorption | [206] |
| SBA-15- | Native one SBA-OH, aminated SBA-$NH_2$, thiolated SBA-SH and carboxy-terminated SBA-COOH. | CD | HSA | SBA-COOH mediated the conformational changes of HSA | [207] |
| AuNPs | D, L, and racemic penicillamine | Quartz microbalance platform | transferrin | Chiral surface of NP dictate the denaturation of transferrin | [208] |
| polystyrene NPs | Poly (phosphoester)s | ITC | HSA | Chemical structure itself plays a crucial role in interaction process. | [209] |
| polystyrene NPs | Sodium dodecyl sulfate or Lutensol AT50 | ITC | HSA | surface chemistry markedly changes the protein binding affinity | [210] |

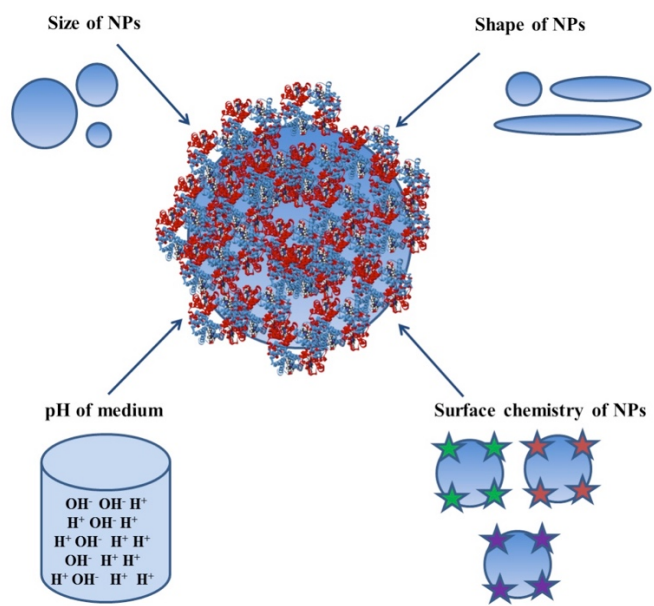

**Figure 1.** Protein corona structure mostly depend on the physicochemical properties of NPs such as size, shape, surface functional groups, and pH of medium.

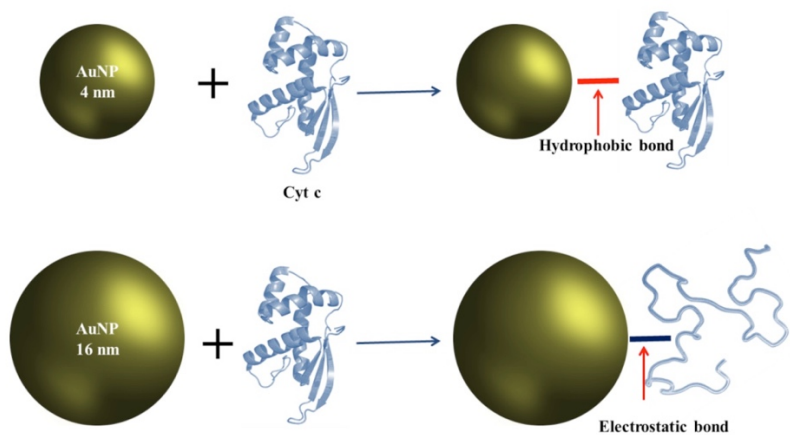

**Figure 2.** Adsorption of protein on larger NPs results in bigger conformational changes compared with interaction with smaller NPs. Hence the size can influence the kind of interactions between NP and proteins.

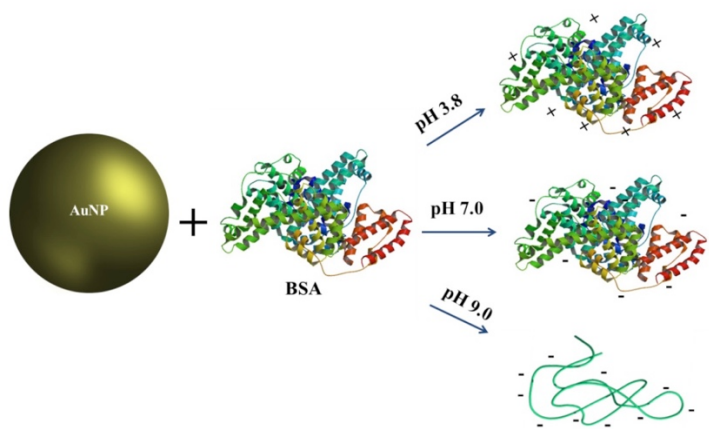

**Figure 3.** The pH-dependent denaturation of BSA after interaction with AuNPs. Different pH of medium induces different charge distribution on the protein surfaces and subsequent interactions and denaturation.

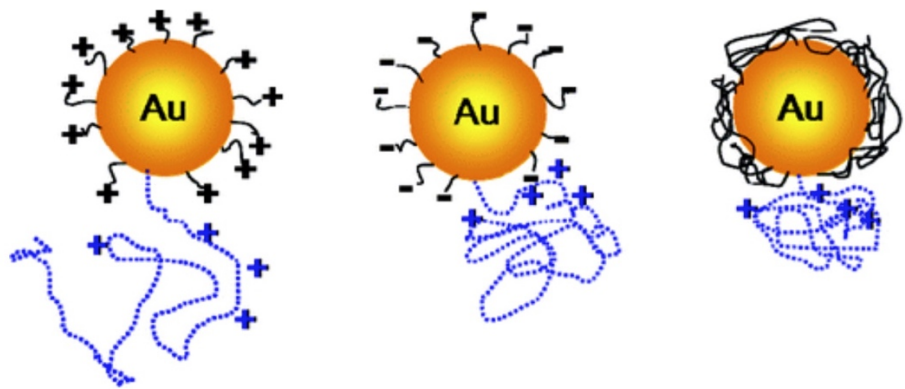

**Figure 4.** The surface chemistry of AuNPs mediates the degree of NP-triggered conformational changes of cyt C. When gold nanoparticles are placed with positive, negative and neutral ligands adjacent to cytochrome C, the structure of cytochrome C changes with the shift in the ions of NPs. Protein maintains its structure with neutral ligands, but changes in the presence of other ions. Reprinted with permission from reference [51].

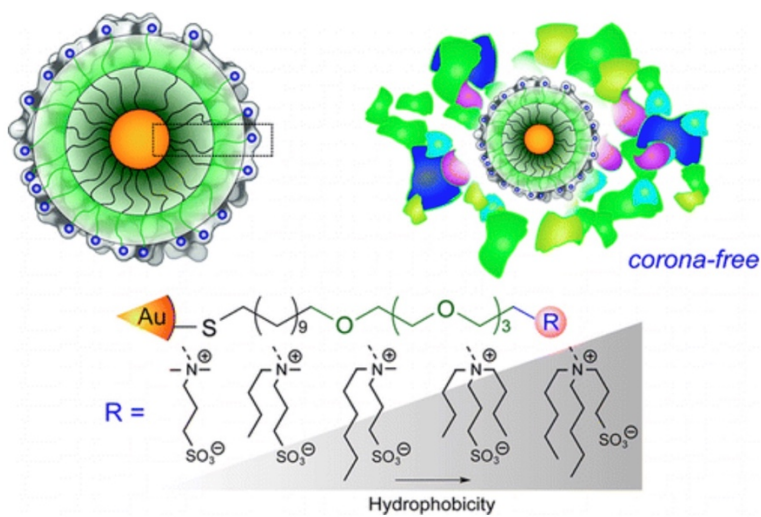

**Figure 5.** Depending on the surface properties of NPs, proteins do not adsorb on NPs. Gold NPs incorporated with a double mixture of hydrophobic and hydrophilic thiolated ligand molecules have been shown to form a ligand shell with stripe-like domains of alternating hydrophobic/hydrophilic combination. These stripe-like domains have a characteristic thickness on the order of a single nanometer, leading to surface heterogeneities at similar scales as those found on proteins. Reprinted with permission from reference [52].

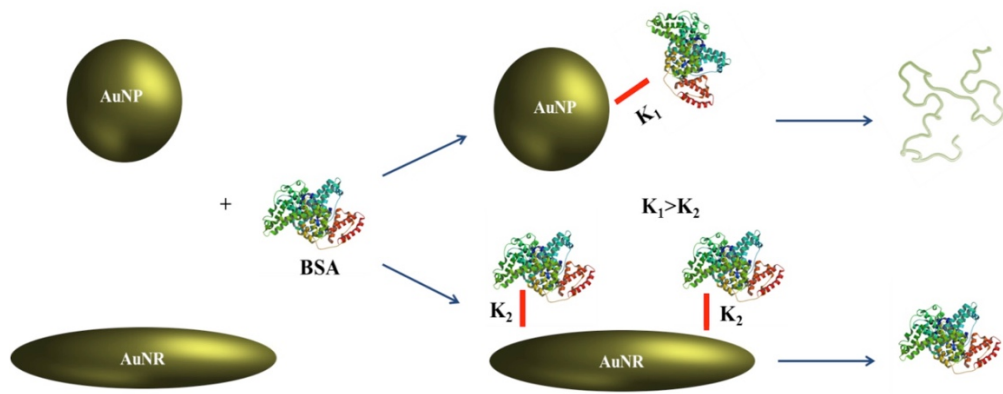

**Figure 6.** Different interactions with Au NPs and corresponding structural changes of BSA dependent on the variations in NP morphology.

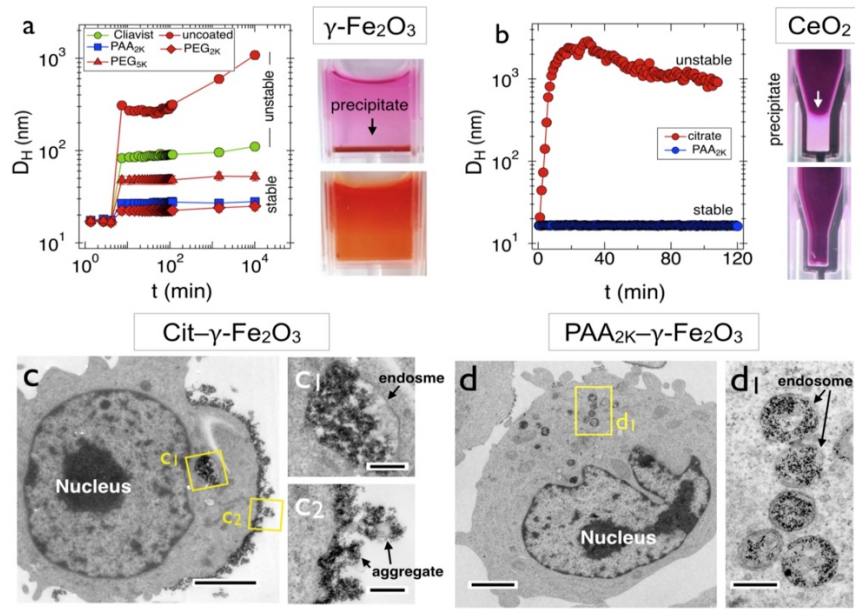

**Figure 7:** a) Hydrodynamic radius $D_H$ *versus* time for 7 nm coated and uncoated iron oxide NPs in the cell culture medium supplemented with 10% heat-inactivated fetal bovine serum. For uncoated particles as well as for cliavist® NP aggregation occurs. Particles coated with polymers (e.g. poly(acrylic acid) and phosphonic acid PEG copolymers) remain disperse for several months and devoid of protein corona [64]. b) Same as in a) for 7 nm cerium oxide NPs ($CeO_2$) coated with citrate and poly(acrylic acid) [84]. c) Transmission electron microscopy (TEM) images of lymphoblastoid cells treated with Cit–$\gamma$-$Fe_2O_3$ NPs for 24 h at [Fe] = 10 mM; c1) and c2) demonstrate NPs enclosed in an endosome and clusters, respectively. Bars are 2 μm in c) and 300 nm in c1-c2) [59]. d) Same as in c) for lymphoblastoid cells incubated with $PAA_{2K}$–$\gamma$-$Fe_2O_3$ NPs for 24 h at [Fe] = 10 mM; b1) exhibits NPs enclosed in 200 nm endosomes. Bars in d) are 2 μm and 300 nm in d1) [59].

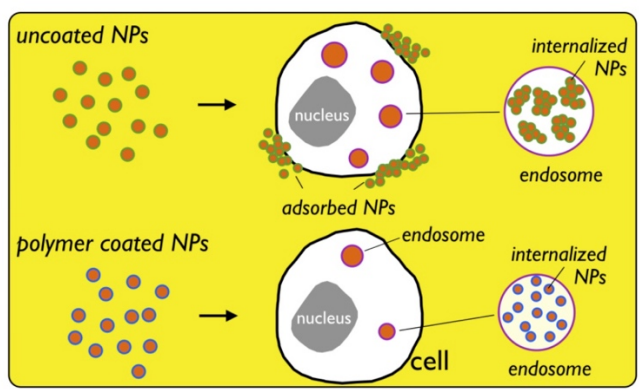

**Figure 8.** Schematic representation of cell-NP interactions. Uncoated NPs aggregate in the presence of serum protein and are either internalized into endosomes in large quantities or sticking at the membrane. Polymer coated NPs in contrast are much less internalized and are located intracellularly in endosomes [59, 84, 211].

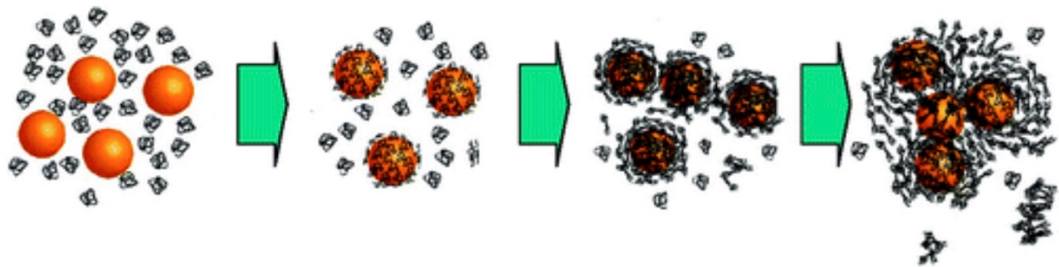

**Figure 9.** AuNPs initiate amyloid nucleation, causing the formation of protein aggregation without embedded NPs. Reprinted with permission from reference [90].

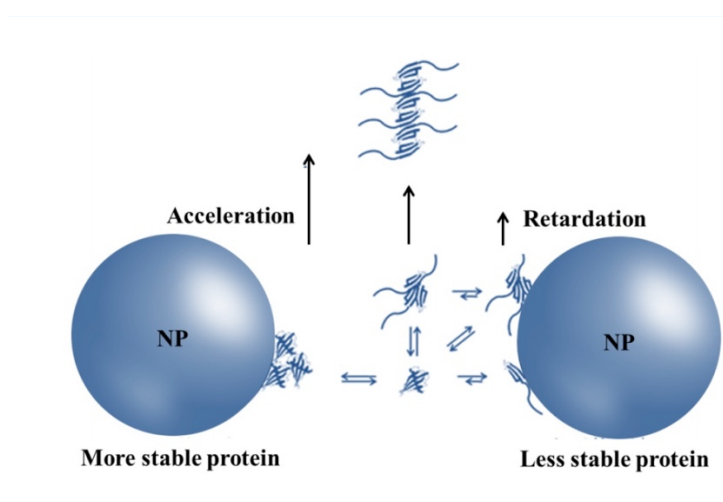

**Figure 10.** NPs accelerated amyloid nucleation of more stable protein

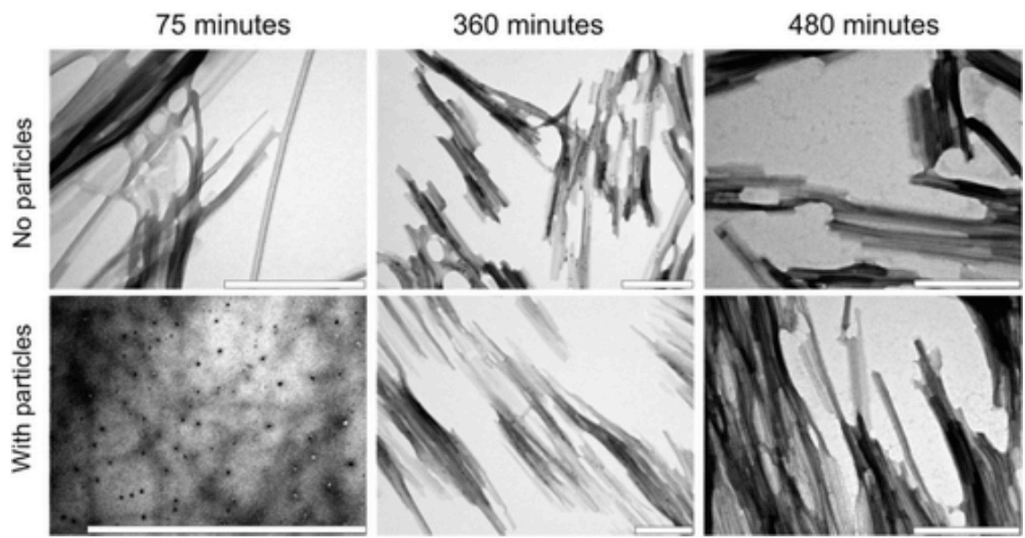

**Figure 11.** Protein adsorption on the NP surface result in small amounts of free protein in the solution. Thus, the binding process causes retardation of amyloid nucleation phase. Reprinted with permission from reference [99].

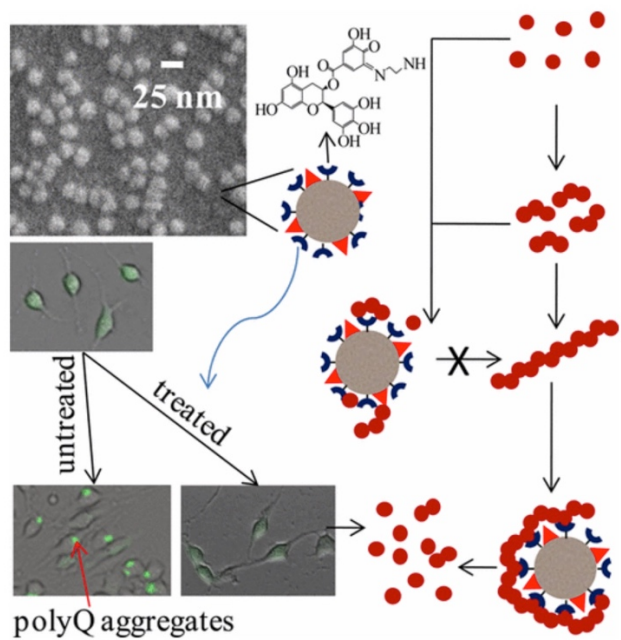

**Figure 12.** Nano-EGCG protects neuronal cells from the adverse effects of extracellular polyglutamine (polyQ) aggregates or protein aggregates by inhibiting their amyloid induction. Reprinted with permission from reference [105].

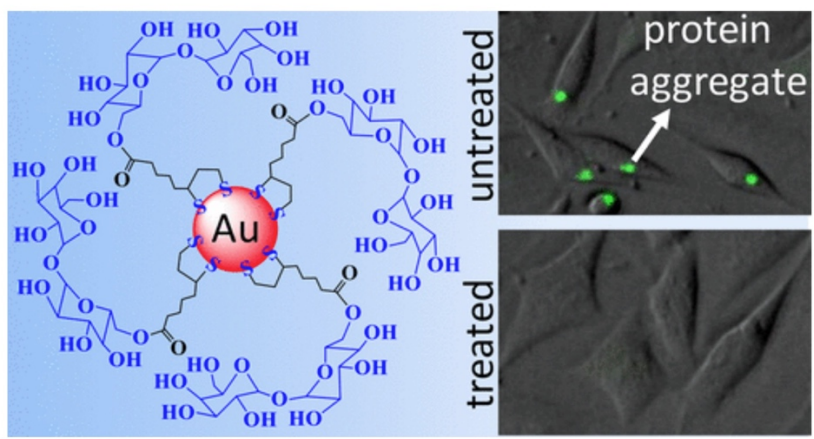

**Figure 13.** Trehalose-modified AuNPs can play an inhibitory impact on intracellular polyglutamine-containing mutant protein amyloid elongation. Reprinted with permission from reference [106].

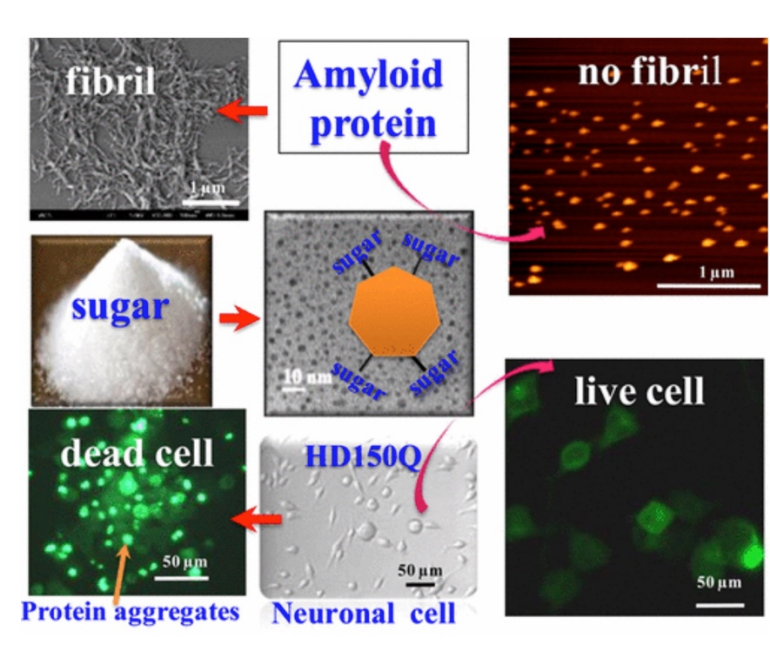

**Figure 14.** Sugar-modified NP inhibits the induction of protein amyloid and decrease the adverse effects of aggregated proteins on the cells. Reprinted with permission from reference [107].

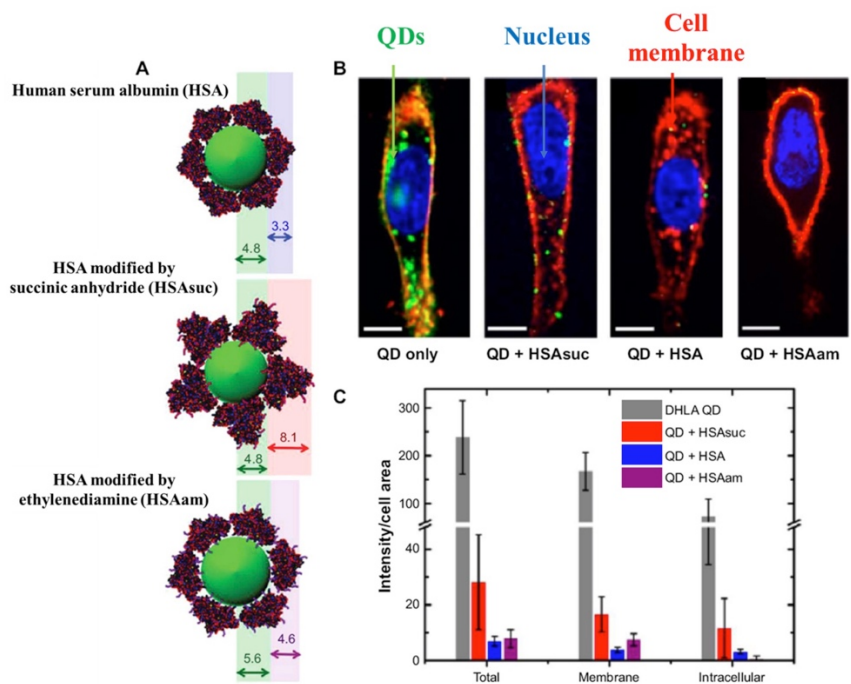

**Figure 15.** Adsorption of HSA on dihydrolipoic acid-coated quantum dots (DHLA-QDs) affected adhesion velocity to the cell membrane and infiltrate outcome for HeLa cells. (A) Adsorption of different HSA proteins onto DHLA-QDs. (B) Cells were treated with QDs in the absence or presence of different HSA proteins. (C) Quantification of NP uptake. Reprinted with permission from reference [122].

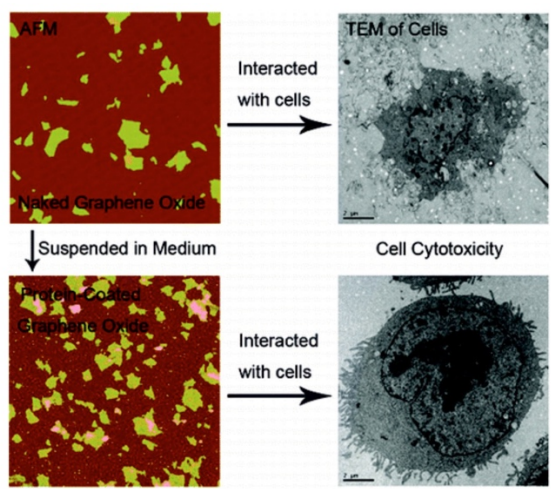

**Figure 16.** The cytotoxicity of graphene oxide was markedly reduced at cell culture medium. Reprinted with permission from reference [125].

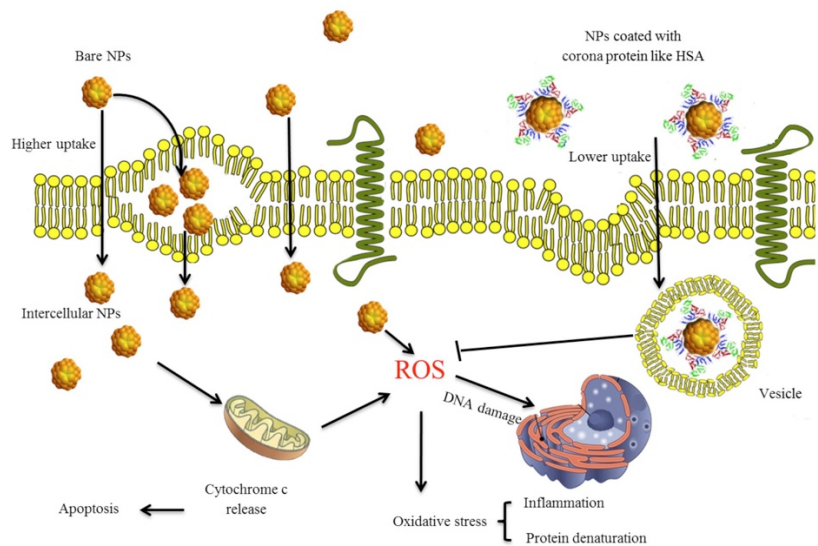

**Figure 17.** Possible uptake process and mechanisms of cytotoxicity induced by NPs and mitigated toxicity by corona protein in cells based on the metadata from several studies.

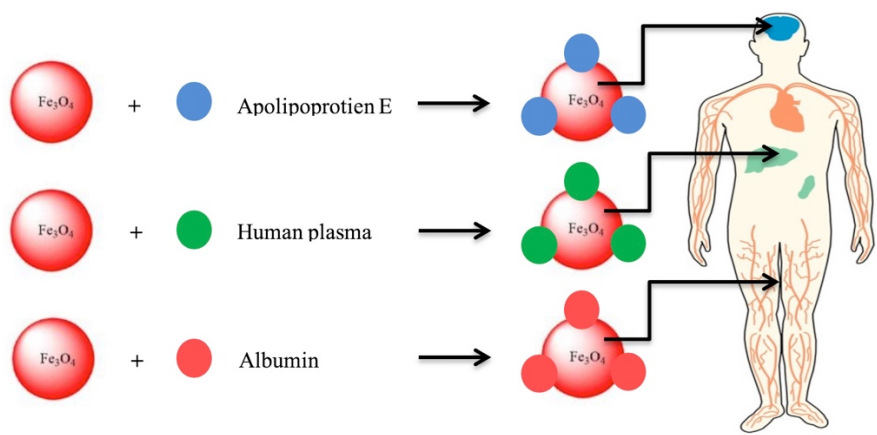

**Figure 18.** Schematic illustration of NPs bio-diffusion with varying coatings.

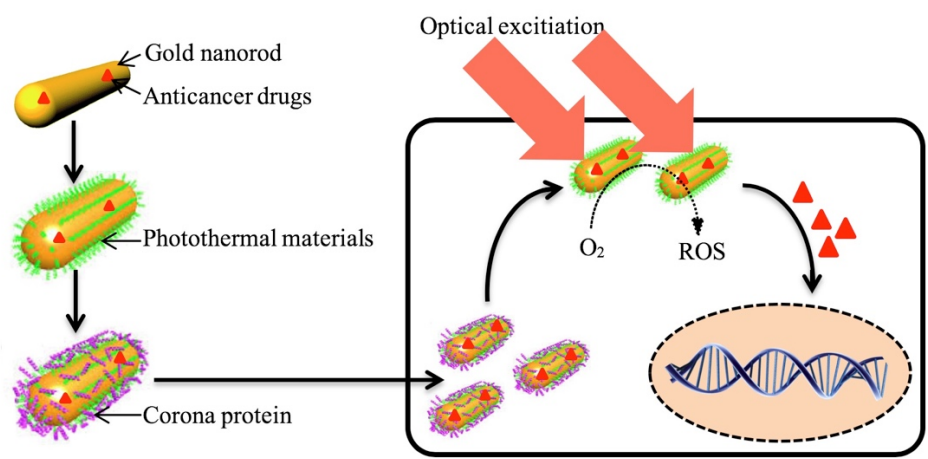

**Figure 19.** Schematic representation of NPs construction and drug transfer with irradiation therapy in target cells.

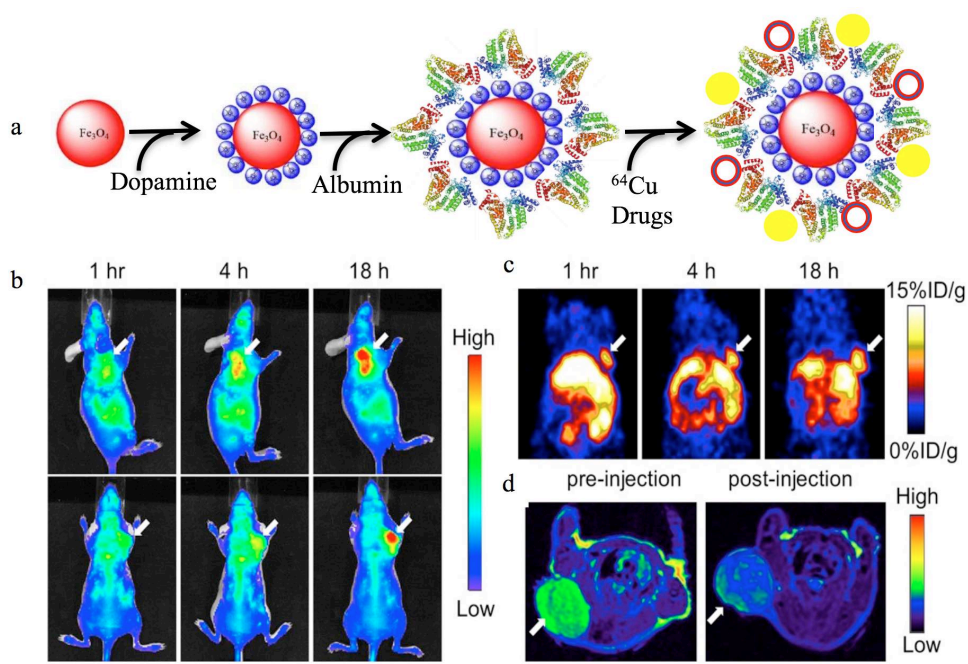

**Figure 20.** (a) Schematic illustration of the development of HSA-IO NPs. (b) Illustrative in vivo NIRF images of mouse inoculated with HSA-IONPs. Images were acquired 1 h, 4 h and 18 h post injection. (c) In vivo PET imaging outcomes of mouse inoculated with HSA-IONPs. Images were obtained 1 h, 4 h and 18 h post injection. (d) MRI images obtained before and 18 h post inoculation. Reprinted with permission from reference [157].